%% file: main.tex
\documentclass[10pt,conference]{IEEEtran}
\usepackage{subfigure}
\usepackage{amsmath,amsfonts}
\usepackage{algorithmic}
\usepackage{algorithm}
\usepackage{array}
\usepackage[caption=false,font=normalsize,labelfont=sf,textfont=sf]{subfig}
\usepackage{textcomp}
\usepackage{stfloats}
\usepackage{url}
\usepackage{verbatim}
\usepackage{graphicx}
\usepackage{cite}
\usepackage{xcolor}
\usepackage{soul}
\usepackage{amssymb}
\usepackage{enumitem}

\usepackage{arydshln}
\usepackage{fancyhdr}
\usepackage{varwidth}
\usepackage{arydshln}
\usepackage{multirow}
\usepackage{mathtools}

\usepackage{datetime}

\usepackage{kotex}
\usepackage{soul}

\newcommand{\sys}{\textsf{\small RT-MOT}}

\newcommand{\eqdef}{\vcentcolon=}

\newcommand{\remove}[1]{}

\newtheorem{lemma}{Lemma}
\newtheorem{theorem}{Theorem}

\long\def\symbolfootnote[#1]#2{\begingroup%
\def\thefootnote{\fnsymbol{footnote}}\footnotetext[#1]{#2}\endgroup}

\begin{document}
\title{RT-MOT: Confidence-Aware Real-Time Scheduling Framework for Multi-Object Tracking Tasks} 

\author{
\IEEEauthorblockN{
Donghwa Kang\IEEEauthorrefmark{2},
Seunghoon Lee\IEEEauthorrefmark{3},
Hoon Sung Chwa\IEEEauthorrefmark{4},
Seung-Hwan Bae\IEEEauthorrefmark{5},
Chang Mook Kang\IEEEauthorrefmark{6},\\
Jinkyu Lee\IEEEauthorrefmark{3} and
Hyeongboo Baek\IEEEauthorrefmark{2}
}

\IEEEauthorblockA{\small{~}}

\IEEEauthorblockA{\small{Dept. of Computer Science and Engineering, Incheon National University (INU), Republic of Korea\IEEEauthorrefmark{2}}\\
}

\IEEEauthorblockA{\small{Dept. of Computer Science and Engineering, Sungkyunkwan University (SKKU), Republic of Korea\IEEEauthorrefmark{3}}\\
}

\IEEEauthorblockA{\small{Dept. of Electrical Engineering and Computer Science, DGIST, Republic of Korea\IEEEauthorrefmark{4}}\\
}

\IEEEauthorblockA{\small{Dept. of Computer Engineering, Inha University, Republic of Korea\IEEEauthorrefmark{5}}\\
}

\IEEEauthorblockA{\small{Dept. of Electrical Engineering, Incheon National University (INU), Republic of Korea\IEEEauthorrefmark{6}}\\
}

\IEEEauthorblockA{\small{hbbaek@inu.ac.kr}\\
}

\vspace{-0.4cm}
}

\markboth{Journal of \LaTeX\ Class Files,~Vol.~14, No.~8, April~2022}%
{Shell \MakeLowercase{\textit{et al.}}: A Sample Article Using IEEEtran.cls for IEEE Journals}

\IEEEpubid{0000--0000/00\$00.00~\copyright~2021 IEEE}

\maketitle

\symbolfootnote[0]{Hyeongboo Baek is the corresponding author of this paper, and 
Jinkyu Lee is the co-corresponding author.}

\thispagestyle{plain}
\pagestyle{plain}

\input{00abstract}

\input{01introduction}

\input{02motivation}

\input{03system_design}

\input{04reliability}

\input{05scheduling}

\input{06evaluation}

\input{07related_work}

\input{08conclusion}

\section*{Acknowledgement}

This work was supported by the National Research Foundation of Korea (NRF) grant funded by the Korea government (MSIT) (2018R1A5A1060031 (ERC),
NRF-2020R1F1A1071547, NRF-2020R1F1A1076058, NRF-2021R1A2B5B02001758, NRF-2021R1F1A1059277, NRF-2021K2A9A1A01101570, NRF-2022R1A4A3018824, NRF-2022R1C1C1009208).

This work was also supported by Institute of Information \& communications Technology Planning \& Evaluation (IITP) grant (IITP-2022-0-01053, IITP-2022-0-00448, Deep Total Recall: Continual Learning for Human-Like Recall of Artificial Neural Networks, 10\%) funded by the Korea government (MSIT), as well as the DGIST R\&D Program of MSIT (20-CoE-IT-01).

\bibliographystyle{IEEEtran}
\bibliography{main}

\end{document}

%% file: 00abstract.tex
\begin{abstract}

Different from existing 
MOT (Multi-Object Tracking) techniques that usually aim at improving tracking accuracy and average FPS, 
real-time systems such as autonomous vehicles necessitate new requirements of MOT under limited computing resources:
(R1) guarantee of timely execution and (R2) high tracking accuracy.
In this paper, we propose RT-MOT, a novel system design for multiple MOT tasks, which addresses R1 and R2.
Focusing on multiple choices of a workload pair of \textit{detection} and \textit{association},
which are two main components of the tracking-by-detection approach for MOT, we tailor a measure of object confidence for RT-MOT
and develop how to estimate the measure for the next frame of each MOT task.
By utilizing the estimation, we make it possible to predict tracking accuracy variation
according to different workload pairs to be applied to the next frame of an MOT task.
Next, we develop a novel \textit{confidence-aware} \textit{real-time} scheduling framework,
which offers an offline timing guarantee for a set of MOT tasks based on non-preemptive fixed-priority scheduling with the smallest workload pair.
At run-time, 
the framework checks the feasibility of a priority-inversion associated with a larger workload pair, which does not compromise the timing guarantee of every task, and
then chooses a feasible scenario that yields the largest tracking accuracy improvement 
based on the proposed prediction.
Our experiment results demonstrate that RT-MOT significantly improves overall tracking accuracy by up to 1.5$\times$, compared to existing popular tracking-by-detection approaches, while guaranteeing timely execution of all MOT tasks.
\end{abstract}

%% file: 01introduction.tex
\section{Introduction}

As Multi-Object Tracking (MOT) is essential for various vision applications,
there have been many attempts to improve tracking accuracy and average FPS for MOT.
In the case of a system integrated with control, if object information is not transmitted according to the control sample period, the controller performs control based on previous data; in the worst case, it can lead to a severe accident in a system such as an autonomous emergency braking system.
Therefore, real-time systems such as autonomous vehicles
necessitate new requirements of MOT under limited computing resources:
(R1) guarantee of timely execution and (R2) high tracking accuracy.
\noindent Although a recent study in~\cite{self-cueing} has addressed both R1 and R2 for a single MOT task,
no study has achieved both for multiple MOT tasks to be applied to a vision system with multiple cameras.

In this paper, we propose \sys{}, a novel system design for 
multiple MOT tasks, which achieves R1 and R2. To this end, we first address the following question.
\begin{itemize}
    \item [Q1.] How can we design the system architecture of \sys{} to provide a control knob to explore a trade-off between R1 and R2?
\end{itemize}
To answer Q1, we focus on 
the tracking-by-detection structure (one of the most popular structures for MOT) 
consisting of 
\textit{detection} of objects in a single video frame and \textit{association} to match objects detected in the current frame with those from previous frames.
Based on the structure, 
\sys{} implements a new tracking-by-detection structure with multiple choices of a pair of detection and association models.
Since different choices for each instance of an MOT task result in different execution times (affecting R1) and different confidence of detected/associated objects
(affecting R2),
the proposed system architecture addresses Q1,
which brings the next question.

\begin{itemize}
    \item [Q2.] How can we efficiently utilize the proposed system architecture to achieve R1 and R2?
\end{itemize}
To answer Q2, we need to investigate how R1 and R2 are affected by 
different pairs of detection and association models selected by an instance of MOT.
To address the R2 part,
we focus on the reliability of detected/associated objects,
and re-define a notion of object confidence for RT-MOT,
consisting of motion confidence and appearance confidence,
updated by the detection and association parts, respectively.
We then develop how to estimate object confidence for the next frame of each MOT,
which enables to predict tracking accuracy variation according to different pairs of detection and association models to be applied to the next framework of an MOT task.
The prediction is key % a key to explore 인것 같은데, 문법 잘 모르겠음
to exploring a trade-off between R1 and R2 to be utilized in the scheduling framework.

We then develop a novel \textit{confidence-aware real-time} scheduling framework designed for \sys{}. To this end, we propose a new scheduling algorithm, 
\mbox{NPFP$^\textsf{flex}$} (Non-Preemptive Fixed-Priority with \textsf{flex}ible execution),
which addresses the R1 and R2 parts of Q2, respectively by offline/online timing guarantee 
and the estimation of object confidence.
\mbox{NPFP$^\textsf{flex}$} has the following salient features.

\begin{itemize}
    \item \mbox{NPFP$^\textsf{flex}$} offers an offline guarantee of timely execution for all instances of a set of MOT tasks scheduled by non-preemptive fixed-priority scheduling,
    assuming every instance executes with a pair of detection and association models that requires the shortest execution time. This achieves R1.
    \item At run-time, \mbox{NPFP$^\textsf{flex}$} checks the feasibility of execution of another instance (by priority-inversion) with another pair (that may require a longer execution time), without compromising the timely execution of all other instances.
    Among all feasible scenarios (i.e., which instance and how long to be executed), 
    \mbox{NPFP$^\textsf{flex}$} chooses the one that yields the largest expected improvement of tracking accuracy based on the estimation of object confidence.
    This achieves R2 without compromising R1.
\end{itemize}

We implemented \sys{} and evaluated its effectiveness in achieving R2 without compromising R1.
Our evaluation results show \sys{} to improve tracking accuracy by up to 1.5$\times$ compared to existing popular tracking-by-detection approaches without violating any timing constraint.
In addition, we demonstrate that \sys{} properly selects a pair of detection and association models frame-by-frame, achieving nearly maximum tracking accuracy (achievable without timing constraints)
with less total computation.

In summary, this paper makes the following contributions.

\begin{itemize}
    \item We motivate the importance of choosing a proper pair of detection and association models to explore a trade-off between R1 and R2
    (Sec.~\ref{sec:motivation}).
    \item We propose the first system design \sys{}, which addresses R1 and R2 for multiple MOT tasks  (outlined in Sec.~\ref{sec:system_design}).
    \item We re-define and estimate a measure, which enables to predict tracking accuracy variation to be used in the scheduling framework
    (Sec.~\ref{sec:confidence}).
    \item We develop a novel confidence-aware real-time scheduling framework for \sys{}, which offers both offline and online timing guarantees with flexible execution
    (Sec.~\ref{sec:scheduling}).
    \item We demonstrate the effectiveness of \sys{} through experiment on an actual computing system (Sec.~\ref{sec:evaluation}).
\end{itemize}

%% file: 02motivation.tex
\remove{
\begin{figure}[t!]
	\centering 
	\includegraphics[width=0.8\linewidth]{figures/background.pdf}
	
	\caption{Tracking-by-detection} \hbcmt{motion과 appearance state이 쓰이는 것이 직관적으로 보이게 수정필요.}
	\label{fig:background}
\end{figure}
}

\section{Target System and Motivation}
\label{sec:motivation}

This section explains our target system and makes key observations by a measurement-based case study that underlies the design policy of RT-MOT. 

\subsection{Target System: DNN-based Multi-Object Tracking}
\label{subsec:background}

An autonomous vehicle is typically equipped with multiple cameras (e.g., front/side/rear cameras) to sense ambient environments. 
Each multi-object tracking task is implemented as a \emph{periodic task} that takes a video frame from a camera and performs DNN-based computation to predict trajectories of multiple targets in frame sequences; this process is repeated periodically. 
Timely response with high tracking accuracy is a necessity because each task has a deadline, and its result is used as input for other components (i.e., motion planning) of the car. 

For multi-object tracking, the tracking-by-detection approach~\cite{deepsort, sort} is one of the most popular ones, and it is characterized by two main steps: 1) \emph{detection} of objects in a single video frame and 2) \emph{association} to match objects detected in the current frame with those from previous frames. 
Thus, its tracking process can be divided into a front-end detector followed by a back-end tracker.
Since object detection is a standalone step in the tracking process, one benefit is the flexibility to pair different object detection models with different association strategies. 

\subsection{Trade-off between Execution Time and Tracking Accuracy}
\label{subsec:motivation}

Although the tracking-by-detection approach offers high flexibility in utilizing various object detection models and association methods,
it is still very challenging to achieve both accurate tracking and timely response on resource-constrained computing platforms
since the execution time and accuracy often conflict with each other. 
We present a case study to investigate the effect of applying different detection models and association methods on execution time and tracking accuracy and identify the main challenges faced therein.

In the case study, we employ YOLOv5\footnote{https://github.com/ultralytics/yolov5}
with two different inputs: i) processing a full-size frame (referred to as \emph{high-confidence detection}) and ii) processing a partial portion of a frame  (referred to as \emph{low-confidence detection});
the details will be explained in Sec.~\ref{subsec:design}.
We also employ two different popular association strategies: i) performing both feature- and Intersection-over-Union (IoU)-based methods (referred to as \emph{high-confidence association}) and ii) performing the IoU-based method only (referred to as \emph{low-confidence association}).

\begin{figure}[t]
	\centering
	\includegraphics[width=1\linewidth]{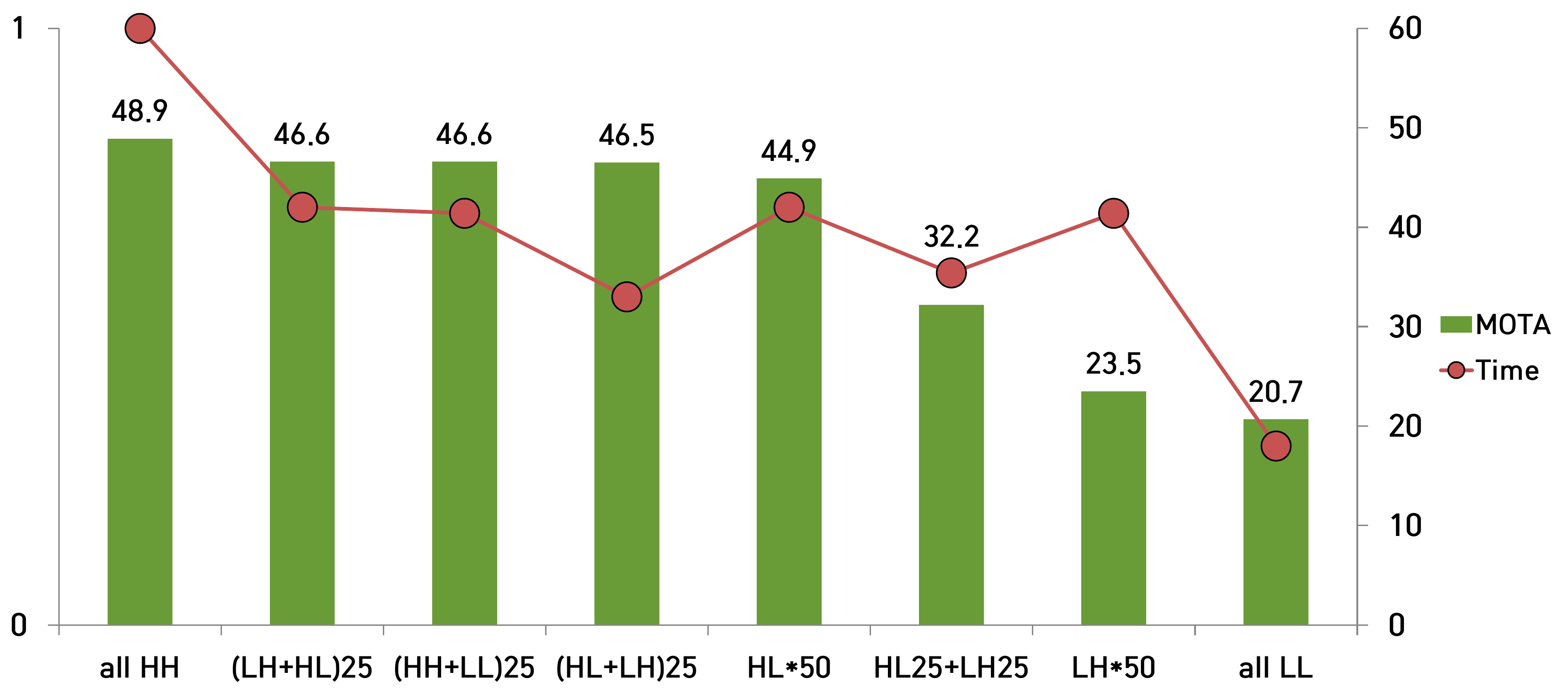}	
	\caption{Tracking accuracy (by a well-known measure MOTA) and execution time (normalized by the largest one) under different combinations of detection and association schemes over 50 consecutive frames}
	\label{fig:motivation}
\end{figure}

Fig.~\ref{fig:motivation} shows the effect of applying 6 different combinations of a pair of detection and association schemes over 50 consecutive video frames on the execution time and tracking accuracy.
In the figure, the x-axis represents different combinations; ($\texttt{D}^x$,$\texttt{A}^y$) expresses a choice of detection and association schemes for a single frame, where $x$ and $y$ can be either \texttt{H} (high-confidence) or \texttt{L} (low-confidence).
For example, (($\texttt{D}^\texttt{H}$,$\texttt{A}^\texttt{H}$)+($\texttt{D}^\texttt{L}$,$\texttt{A}^\texttt{L}$))$\cdot 25$ denoted by (\texttt{H}\texttt{H}+\texttt{L}\texttt{L})$\cdot 25$ refers 25 repetitions of two consecutive frames, which perform a combination of high-confidence detection and high-confidence association, and then 
that of low-confidence detection and low-confidence association.
We consider meaningful combinations whose \texttt{H} (as well as \texttt{L}) ratio for detection/association is exactly 50\%, including 50 identical frames, 25 identical chunks each consisting of two different frames, and 25 identical frames followed by another 25 identical frames. In addition, we add all \texttt{H}\texttt{H} and all \texttt{L}\texttt{L} for reference.

We summarize the following observations from Fig.~\ref{fig:motivation}. 

\begin{itemize}
    \item[\textbf{O1.}] There exists a trade-off between execution time and tracking accuracy in choosing the ratio of high-confidence detection/association. 
\end{itemize}
A higher ratio of high-confidence detection/association achieves higher tracking accuracy at the expense of more execution time. 
For example, (\texttt{H}\texttt{H})
improves the tracking accuracy by 48.9\% but requires more execution time by $3.3\times$ over (\texttt{L}\texttt{L}).
Note that we use MOTA (Multiple Object Tracking Accuracy), one of the widely used object tracking accuracy metrics, for accuracy measurement, which will be discussed in Sec.~\ref{sec:evaluation}.
Also, note that the execution time shown in Fig.~\ref{fig:motivation} is the total sum of the execution times of all frames and is normalized to the case of (\texttt{H}\texttt{H}).

\begin{itemize}
    \item[\textbf{O2.}] Tracking accuracy varies greatly with different combinations of a choice of detection and association schemes, although the combinations yield similar computation time.
\end{itemize}
In Fig.~\ref{fig:motivation}, all combinations except 
(\texttt{H}\texttt{H}) and (\texttt{L}\texttt{L})
have the same ratio (i.e., 50\%) of high-confidence detection and association applied to 50 consecutive video frames, resulting in comparable execution times.
On the other hand, tracking accuracy varies widely up to a difference of
14.4
percentage points (\%p) depending on not only the choice of detection and association schemes for each frame but also the collection of choices for consecutive frames. 
For example, (\texttt{H}\texttt{H}+\texttt{L}\texttt{L})$\cdot 25$ and 
\texttt{H}\texttt{L}$\cdot 25$+\texttt{L}\texttt{H}$\cdot 25$ shows 46.6\%
and 32.2\%
of tracking accuracy, respectively (meaning that the former yields 44.7\% higher accuracy than the former), while their total computation times differ only by 
 16.4\%.
 
Observation O1 and O2 give rise to % both 
opportunities and challenges \textit{together} in achieving
both timely response and maximum tracking accuracy under limited computing resources.
By O1, we can \emph{dynamically} decide which detection and association schemes are to be applied for each frame of a multi-object tracking task by trading off the execution cost for accuracy.
If there is less spare time before the next frame arrives, a multi-object tracking task can select low-confidence detection and/or association to finish its execution before the deadline at the expense of sacrificing tracking accuracy.
On the other hand, if there is enough spare time before the next frame arrives, a multi-object tracking task can select both high-confidence detection and association to improve tracking accuracy.

By O2, it is nevertheless difficult to find an optimal combination of a choice of detection and association schemes for each task that maximizes tracking accuracy without compromising the timely execution of the task. 
This is because i) there exists a huge number of possible combinations of the choices of detection and association schemes for a frame sequence, and ii) the tracking accuracy of one combination also dynamically varies with the input frame sequence. 
Moreover, when multiple tracking tasks are scheduled together by sharing limited computing
resources,
a selected combination of detection and association schemes for a task
can affect not only the timely execution and tracking accuracy of the task itself,
but also those of the other tasks. This makes it very challenging to find the right solution.

\remove{
\begin{figure}[t!]
	\centering 
	\includegraphics[width=1\linewidth]{figures/motivation.jpg}
	
	\caption{Overall system design}
	\label{fig:design_new}
\end{figure}
}

\begin{figure*}[t!]
	\centering 
	\includegraphics[width=\linewidth]{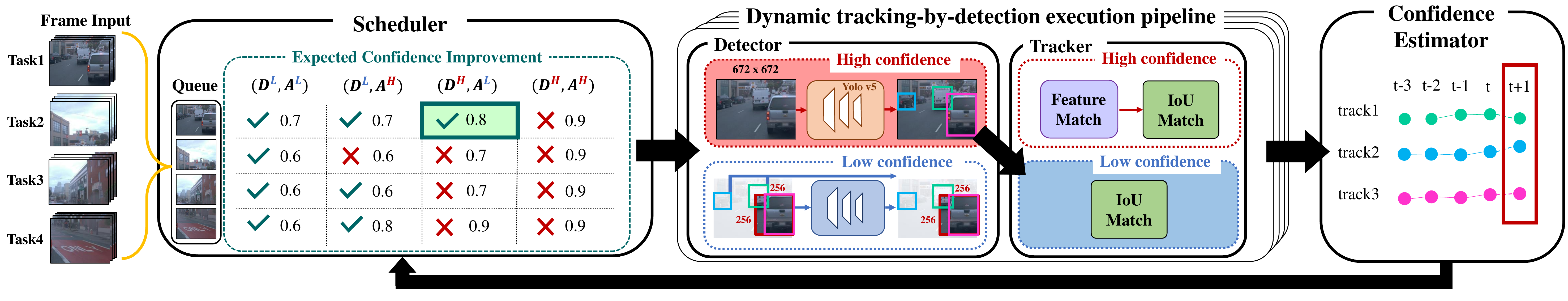}
	\caption{Overall system design of \sys{}}
	\label{fig:design_new}
\end{figure*}

%% file: 03system_design.tex
\section{\sys{}: System Design}
\label{sec:system_design}

In this section,
we present our goal and design of \sys,
based on O1 and O2 in Sec.~\ref{sec:motivation}.

\subsection{System Goal and Overview}

We aim to develop an online tracking-accuracy-aware scheduling framework, which achieves
the following goals for real-time multi-object tracking tasks:

\begin{itemize}
    \item[\textbf{R1.}] It provides timing guarantees for real-time multi-object tracking tasks; and
    \item[\textbf{R2.}] It maximizes overall tracking accuracy. 
\end{itemize}

To this end, we propose a new system abstraction, named \sys{}, that enables frame-level flexible scheduling of multi-object tracking tasks.
In particular, \sys{} supports dynamic selection of different execution models for detection and association, respectively, for each frame; it %and 
makes run-time frame-level scheduling decisions by considering the effect of execution model selection on both timely execution and tracking accuracy of multi-object tracking tasks. 
We address the following key issues, which enables \sys{} to maximize overall tracking accuracy while meeting all timing constraints:

\begin{itemize}
    \item[\textbf{I1.}] How to estimate the variation of overall accuracy according to different detector/tracker selections of the next frame?
    \item[\textbf{I2.}] How to provide an offline timing guarantee to multiple MOT tasks while 
    maximizing the overall accuracy at run-time using the answer of I1?
    \item[\textbf{I3.}] How to design the system architecture that supports flexible tracking-by-detection and provides an interface that can accommodate the answer of I1 and I2?
\end{itemize}

\subsection{Design of \sys{}}
\label{subsec:design}

\sys{} supports frame-level flexible scheduling for multi-object tracking tasks to maximize overall tracking accuracy while providing real-time guarantees. 
As depicted in Fig.~\ref{fig:design_new},
the core design features of \sys{} include dynamic tracking-by-detection execution pipeline (addressing I3), multi-object tracking confidence estimator (addressing I1 and to be detailed in Sec.~\ref{sec:confidence}), and frame-level flexible scheduler (addressing I2 and to be detailed in Sec.~\ref{sec:scheduling}) as follows.

\textbf{Dynamic tracking-by-detection execution pipeline.} \sys{} implements a new tracking-by-detection approach that is capable of combining different front-end detectors with different back-end trackers at frame by frame, each of which exhibits different execution costs and tracking confidence. 

Under \sys{}, each input frame of a multi-object tracking task is forwarded to its corresponding tracking-by-detection execution pipeline. 
The front-end detector identifies the location and size of each object's bounding box in the input frame and sends the detection information to the back-end tracker. 
The back-end tracker then associates each detected object with one of the existing tracks from previous frames (i.e., tracklets) based on motion and/or appearance similarity
and updates the tracking information of the matched tracklets.

We consider two types of execution model for detection and association, respectively: \emph{high-confidence} and \emph{low-confidence} execution models as shown in Fig.~\ref{fig:design_new}. 
The high-confidence execution models exhibit higher tracking confidence at the expense of more execution time, while the low-confidence execution models consume less execution time at the expense of sacrificing tracking confidence.

The tracking-by-detection execution pipeline uses YOLOv5 as a front-end detector.
It can accept variable frame sizes as input, and its inference latency depends on the input size. 
The front-end detector employs YOLOv5 with two different inputs: i) processing an entire frame with the size of $672\times672$ (referred to as \emph{high-confidence detection}) and ii) processing a partial portion of a frame only containing objects of interest with the size of $256\times256$ (referred to as \emph{low-confidence detection}).
For low-confidence detection, we divide an input frame into smaller portions with the size of $256\times256$, extract a particular frame portion by image cropping, and use the cropped image as input for YOLOv5.
Processing a cropped portion with a smaller size can effectively lower the computational workload with no detection accuracy drop in the cropped portion~\cite{KCK22}.
Note that the particular portion of a frame will be henceforth referred to as \emph{Region-of-Interest} (RoI).
Among the frame portions as large as RoI,
we select the one whose average confidence score of the tracklets therein is the lowest,
which efficiently improves tracking accuracy, where confidence score will be explained in Sec.~\ref{sec:confidence}.
To handle the rest portions outside of RoI, a set of tracks outside of RoI (obtained by the previous frame) is also included in the result of detection to keep maintained by the back-end tracker.

The back-end tracker adopts two popular association methods: i) performing both feature- and Intersection-over-Union (IoU)-based methods (referred to as \emph{high-confidence association}) and ii) performing the IoU-based method only (referred to as \emph{low-confidence association}). 
The IoU-based method~\cite{sort} is known as the simplest form of object association.  
The feature-based method~\cite{deepsort} is one of the advanced association methods to handle tracking loss due to occlusion between objects. 
It incorporates a re-identification model as a feature extractor to extract the feature vectors of the detected objects and uses the feature vectors to match already confirmed targets against new detected objects to re-identify occluded targets that are temporally lost, which involves extra computational cost.
Therefore, our proposed tracking-by-detection execution pipeline has four ($2 \times 2$) different choices of a pair of detection and association models in total; each pair is denoted as ($\texttt{D}^x$,$\texttt{A}^y$), where $x$ and $y$ can be either \texttt{H} (high-confidence) or \texttt{L} (low-confidence).
Among them, the scheduler dynamically chooses one pair to be processed for each frame of a multi-object tracking task. 

\textbf{Multi-object tracking confidence estimator.} We define a novel multi-object tracking confidence metric to gauge the confidence level of a list of tracklets for each multi-object tracking task. 
Using this metric, the confidence estimator estimates how the confidence level will vary depending on the choice of a pair of detection and association models to process the next frame of a task. 

After each task finishes its tracking-by-detection execution pipeline, the confidence estimator evaluates the confidence of each tracklet in consideration of the length and continuity of a tracklet and the similarity with the associated detection (see Sec.~\ref{subsec:confidence} for details).
In general, a tracklet has a higher confidence score if it is frequently and recently associated with a detected object having a strong affinity.
The confidence estimator then predicts the amount of potential increase
in each tracklet's confidence score depending on the choices of a pair of detection and association models for the next frame of a task (see Sec.~\ref{subsec:estimation} for details). 
Such a predicted decrease in the confidence score is utilized by the scheduler to determine a pair of detection and association models for each frame and a schedule for tasks.

\textbf{Frame-level flexible scheduler.} 
\sys{} implements an online scheduler that not only dynamically determines a pair of detector and tracker for each frame but also adaptively generates a flexible schedule by considering confidence estimates for tasks to maximize overall tracking accuracy without violating any timing constraints at run-time. 

The scheduler runs as a background daemon to communicate with the tracking-by-detection execution pipelines through the IPC-based communication stub interface implemented within each task and the scheduler.
The scheduler is invoked upon arrival of a new frame for each task, or completion of a task. 
Based on each task's static model parameters (in Sec.~\ref{subsec:model}),
the scheduler determines each frame's priority and its pair of detector and tracker according to non-preemptive fixed-priority with the minimum execution by default (in Sec.~\ref{subsec:NPFPmin}).
However, upon each invocation,
the scheduler checks the possibility of execution of a frame 
(regardless of its priority) with a pair of detector and tracker that yields the maximum overall tracking accuracy without violating any timing constraints guaranteed by the offline schedulability analysis (in Sec.~\ref{subsec:NPFPflex}).

%% file: 04reliability.tex
\section{Tracking Confidence Estimation}
\label{sec:confidence}

This section describes the notion of tracklet confidence tailored to the proposed dynamic tracking-by-detection execution pipeline and 
presents a tracklet confidence estimation method that plays a key role in pair selection and scheduling decisions by the scheduling framework of \sys{}.

\subsection{Tracklet Confidence}\label{subsec:confidence}

We employ the notion of \emph{tracklet confidence} to measure the reliability of a tracklet constructed during object tracking over time, which is widely used in the field of multi-object tracking;
a tracklet is generated by the frame-by-frame association based on the tracking-by-detection approach~\cite{Voigtlaender_CVPR19, Wang_ECCV20}.
We tailor the notion of tracklet confidence to the proposed dynamic tracking-by-detection execution pipeline and explain how to calculate the tracklet confidence score.  

For a video frame, let $O_i^t$ denote as the $i$-th object response detected at the $t$-th frame, and it is characterized by ($M_i^t$, $A_i^t$), where $M_i^t$ and $A_i^t$ are the \emph{motion} and \emph{appearance} states of $O_i^t$, respectively. 
The motion state $M_i^t$ is further specified as $M_i^t = (p_i^t,s_i^t,v_i^t)$, where $p_i^t = (x_i^t, y_i^t)$, $s_i^t = (w_i^t,h_i^t)$, $v_i^t = (vx_i^t, vy_i^t, vw_i^t, vh_i^t)$ are the position, size, and velocity of $O_i^t$, respectively. 
The appearance state $A_i^t$ is the feature vector of $O_i^t$ extracted by a feature extractor. 
Note that $M_i^t$ and $A_i^t$ can be obtained from the results of the front-end detector and back-end tracker, respectively, in the tracking-by-detection execution pipeline. 
We then define a tracklet $\chi_i^t$ of object $O_i^t$ as a set of tracks followed by $O_i^t$ up to the $t$-th frame, and it is expressed as $\chi_i^t = \{O_i^k | 1 \leq t_i^s \leq k \leq t_i^e \leq t \}$, where $t_i^s$ and $t_i^e$ are the start- and end-frame of the tracklet.
A set of tracklets of all objects up to the $t$-th frame is denoted as $\Phi_{1:t}$.

We now explain how to calculate the tracklet confidence score.
A tracklet with a high confidence score is considered as a reliable tracklet, while another tracklet with a low confidence score is considered as an unreliable tracklet with a fragmented trajectory due to inaccurate detection and/or association. 
We model tracklet confidence $\Omega(\chi_i^t)$ of $\chi_i^t$ as $\Omega(\chi_i^t) = \Omega_M(M_i^t) \times \Omega_A(A_i^t)$, where $\Omega_M(M_i^t)$ and $\Omega_A(A_i^t)$ are motion and appearance confidence of $\chi_i^t$, respectively, each of whose range is [0,1]. 

Under \sys{}, after each task finishes its tracking-by-detection execution pipeline for the $t$-th frame, 
the states of each tracklet $\chi_i^t \in \Phi_{1:t}$ are %is 
updated as well as its confidence score.
Depending on the result of the association between a set of tracklets and a set of detected objects at the $t$-th frame, 
each tracklet $\chi_i^t \in \Phi_{1:t}$ can be classified into three categories:

\begin{itemize}[leftmargin=40pt]
    \item [CG1.] $\chi_i^t$ is matched with one of the detected objects by \emph{high-confidence} association; 
    \item [CG2.] $\chi_i^t$ is matched with one of the detected objects by \emph{low-confidence} association; and
    \item [CG3.] $\chi_i^t$ is unmatched with any of the detected objects. 
\end{itemize}
Note that the detected objects shown in CG1--CG3 are the results of either high- or low-confidence detection.
Then, the motion and appearance confidence scores of the tracklet $\chi_i^t$, which corresponds to one of the CG1, CG2, and CG3 cases, can be updated as follow: 

\begin{subequations}
\begin{itemize}
    \item $\Omega_M(M_i^t) = \Omega_A(A_i^t) = 1$, 
    \begin{flushright}if $\chi_i^t$ belongs to CG1; \refstepcounter{equation}\textup{(\theequation)} \label{subeq:a}\end{flushright}
    \item $\Omega_M(M_i^t) = 1$ and $\Omega_A(A_i^t) = \big[\Omega_A(A_i^{t\text{-}1}) \times \Delta A_i^{t\text{-}1}\big]_0$,
    \begin{flushright}if $\chi_i^t$ belongs to CG2; \refstepcounter{equation}\textup{(\theequation)} \label{subeq:b}\end{flushright}
    \item $\Omega_M(M_i^t) = \big[\Omega_M(M_i^{t\text{-}1}) \times \Delta M_i^{t\text{-}1}\big]_0$ and $\Omega_A(A_i^t) = \big[\Omega_A(A_i^{t\text{-}1}) \times \Delta A_i^{t\text{-}1}\big]_0$,
    \begin{flushright}if $\chi_i^t$ belongs to CG3. \refstepcounter{equation}\textup{(\theequation)}\label{subeq:c}\end{flushright}
\end{itemize}
\end{subequations}

\noindent
where $[X]_Y \eqdef \max(X,Y)$, and we denote by $\Delta M_i^{t\text{-}1}$ and $\Delta A_i^{t\text{-}1}$ the amounts of variations of the motion and appearance states ($M_i^{t\text{-}1}$ and $A_i^{t\text{-}1}$) at the $(t\text{-}1)$-th frame from the most-recently-updated motion and appearance states before the $(t\text{-}1)$-th frame (denoted as $M_i^{t\text{-}f}$ and $A_i^{t\text{-}g}$), respectively, where $f, g > 1$. 
We can calculate $\Delta M_i^{t\text{-}1}$ as

\noindent
\begin{align}
   \Delta M_i^{t\text{-}1} =  \Lambda_{s}(M_i^{t\text{-}f}, M_i^{t\text{-}1}) \times \Lambda_{v}(M_i^{t\text{-}f}, M_i^{t\text{-}1}),
\end{align}
where

\noindent
\small
\begin{align}
    \Lambda_{s}(M_i^{t\text{-}f}, M_i^{t\text{-}1}) & = -\frac{1}{4} \bigg( \frac{h_i^{t\text{-}f} - h_i^{t\text{-}1}}{h_i^{t\text{-}f}+h_i^{t\text{-}1}} + \frac{w_i^{t\text{-}f} - w_i^{t\text{-}1}}{w_i^{t\text{-}f}+w_i^{t\text{-}1}} \bigg) + \frac{1}{2}, \label{eq:lambda_s}\\
    \Lambda_{v}(M_i^{t\text{-}f}, M_i^{t\text{-}1}) & = 1 -  2 \bigg\vert \sigma (\frac{vx_i^{t\text{-}f} - vx_i^{t\text{-}1}}{vx_i^{t\text{-}f}+vx_i^{t\text{-}1}} + \frac{vy_i^{t\text{-}f} - vy_i^{t\text{-}1}}{vy_i^{t\text{-}f}+vy_i^{t\text{-}1}})  - \frac{1}{2} \bigg\vert,
\end{align}
\normalsize
\noindent
{and} %where 
$\sigma$ is a sigmoid function~\cite{lu2020retinatrack, braso2020learning}.
We also calculate $\Delta A_i^{t\text{-}1}$ as

\noindent
\begin{align}
    \Delta A_i^{t\text{-}1} = \Lambda_{a}(A_i^{t\text{-}g}, A_i^{t\text{-}1}) = \frac{(A_i^{t\text{-}g} \cdot A_i^{t\text{-}1})}{|A_i^{t\text{-}g}| |A_i^{t\text{-}1}|}.\label{eq:lambda_a}
\end{align}

\noindent
The terms $\Lambda_{s}(M_i^{t\text{-}f}, M_i^{t\text{-}1})$, $\Lambda_{v}(M_i^{t\text{-}f}, M_i^{t\text{-}1})$, and $\Lambda_{a}(A_i^{t\text{-}g}, A_i^{t\text{-}1})$ shown in Eqs.~(\ref{eq:lambda_s})--(\ref{eq:lambda_a}) measure the amounts of variations of size, velocity, and appearance, respectively, each of whose range is in [0,1].\footnote{Those terms are widely used in other object tracking papers, e.g., \cite{bae2017confidence}, as a form of exponential functions. For simplicity, we normalize the exponential functions to be in the range of [0,1].}
Intuitively, if the size (\textit{likewise} acceleration/deceleration) of a tracklet becomes smaller (\textit{likewise} larger) from the ($t\text{-}f$)-th to ($t\text{-}1$)-th frame, the value of $\Lambda_{s}(M_i^{t\text{-}f}, M_i^{t\text{-}1})$ (\textit{likewise} $\Lambda_{v}(M_i^{t\text{-}f}, M_i^{t\text{-}1})$) is close to 0, yielding a large decrease in the motion confidence score of the tracklet if unmatched at the $t$-th frame.
%cmt{(larger)에 해당하는것은 거꾸로 바뀌는건가? 서술에 없음.}
Similarly, if there exists a large dissimilarity between the feature vectors of the tracks at the ($t\text{-}g$)-th and ($t\text{-}1$)-th frame due to occlusion, the value of $\Lambda_{a}(A_i^{t\text{-}g}, A_i^{t\text{-}1})$ is close to 0, yielding a large decrease in the appearance confidence score of the tracklet if unmatched at the $t$-th frame. 
For the newly detected objects that are not associated with any of the existing tracklets in $\Phi_{1:t}$, we create a new tracklet for them to track their trajectories from the $t$-th frame.

\begin{figure}[t]
	\centering 
	\includegraphics[width=1\linewidth]{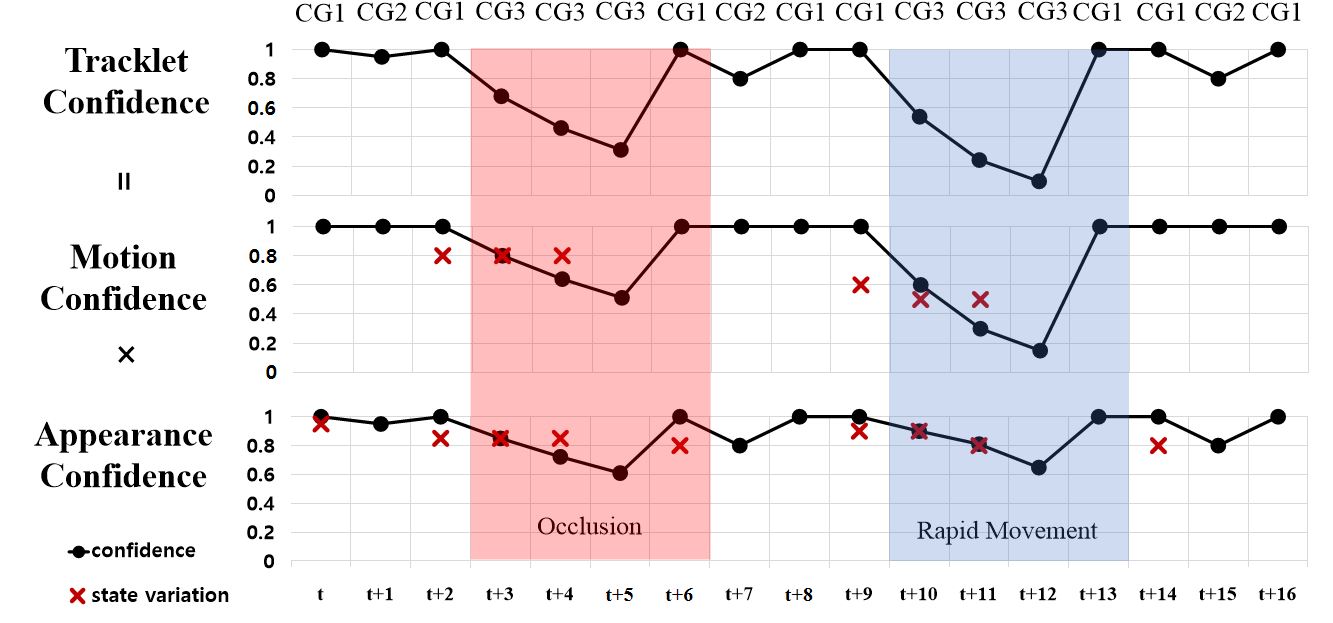}
	
	\caption{{An example of a series of tracklet confidence change according to each frame's situation among CG1--CG3}}
	\label{fig:confidence_example}
\end{figure}

Fig.~\ref{fig:confidence_example} demonstrates how tracklet confidence (accordingly with its motion and appearance confidence) varies depending on CG1--CG3. 
From frame ($t\text{+}3$) to frame ($t\text{+}5$), a target object is temporarily occluded by other objects, so its corresponding tracklet is unmatched (belonging to CG3). 
{Therefore, Eq.~\eqref{subeq:c} updates its motion and appearance confidence scores (represented by black dots in the figure). As seen in Eq.~\eqref{subeq:c},
the motion confidence score of the current frame is calculated by multiplication of that of the previous frame and its state variation (represented by red crosses in the figure);
considering the state variation in $[0,1]$, the motion confidence score of the current frame decreases.
The same holds for the appearance confidence scores as shown in Eq.~\eqref{subeq:c}.
Therefore, the tracklet confidence score (i.e., multiplication of the motion and appearance confidence scores) decreases {from 1 at frame ($t\text{+}2$) to 0.31 at frame ($t\text{+}5$)}.}
In frame ($t\text{+}6$), the occluded object is correctly re-identified by high-confidence association (belonging to CG1).
Then, the tracklet confidence score is set to 1.
A similar trend can be seen from frame ($t\text{+}10$) to frame ($t\text{+}12$) when the target object exhibits rapid movement with high acceleration.
Meanwhile, the tracklet exhibits a large variation in its motion state, yielding a larger decrease in its motion confidence score so does the tracklet confidence score.\\

\subsection{Tracklet Confidence Prediction for the Next Frame}\label{subsec:estimation}

\begin{figure}[t]
	\centering 
	\includegraphics[width=1\linewidth]{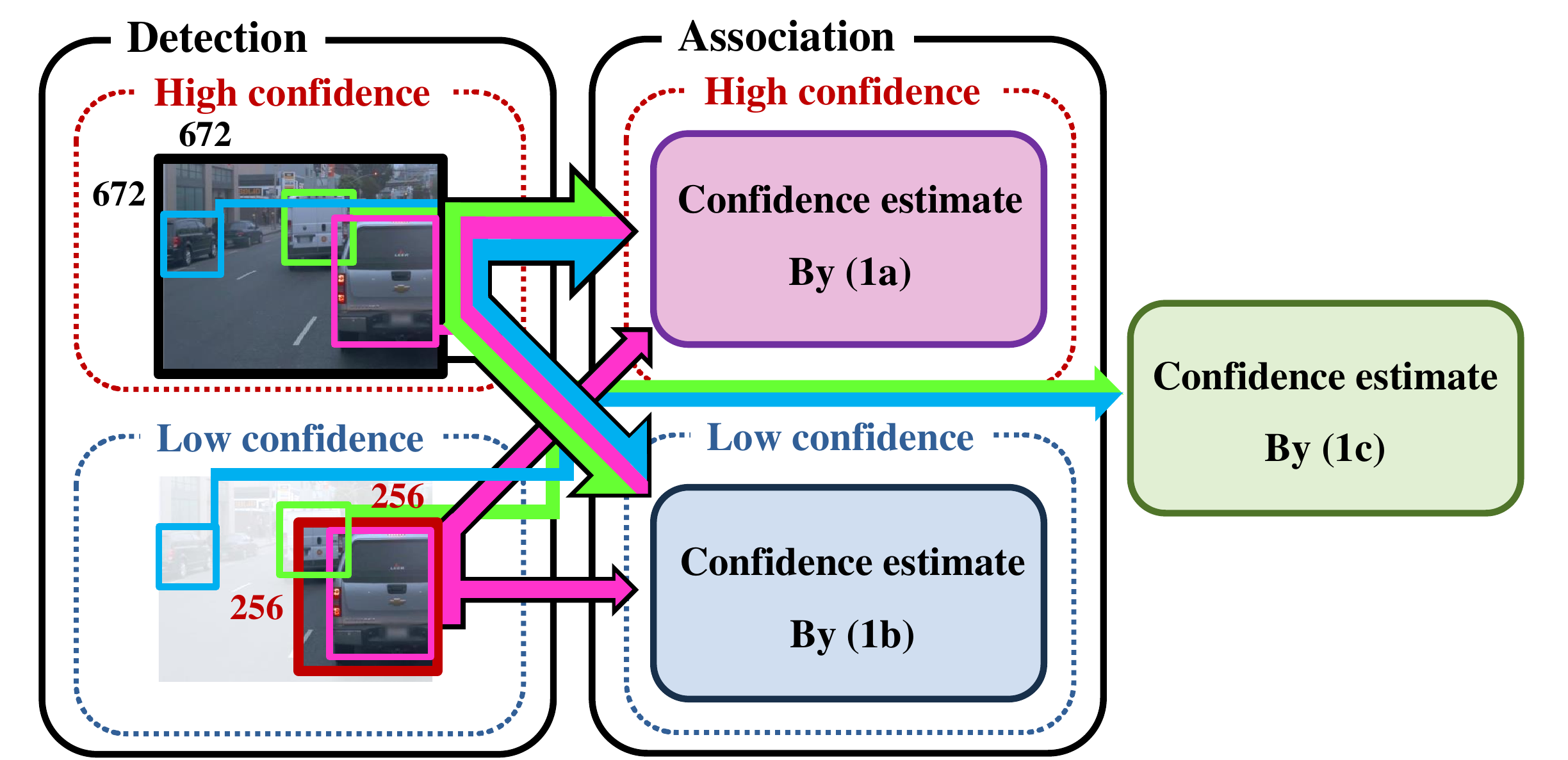}
	\caption{{An illustration of how each tracket calculates its confidence estimate 
	according to different choices of a pair of detection and association models}}
	\label{fig:estimation}
\end{figure}

We now present a method to predict %how to estimate 
tracklet confidence depending on different choices of
a pair of detection and association models for the next frame, which will play a key role in pair selection and scheduling decision by the scheduling framework (to be discussed in Sec.~\ref{sec:scheduling}) to improve overall tracking accuracy. 
For a given set of tracklets $\Phi_{1:t}$ up to the $t$-th frame, 
we construct $\Phi_{1:t\text{+}1}$ and 
calculate a tracklet confidence estimate $\overline{\Omega}(\chi_i^{t\text{+}1})$ at the $(t\text{+}1)$-th frame based on the following assumptions, for four cases of ($\texttt{D}^\texttt{H}$,$\texttt{A}^\texttt{H}$), ($\texttt{D}^\texttt{L}$,$\texttt{A}^\texttt{H}$), ($\texttt{D}^\texttt{H}$,$\texttt{A}^\texttt{L}$), and ($\texttt{D}^\texttt{L}$,$\texttt{A}^\texttt{L}$) as shown in Fig.~\ref{fig:estimation}:

{
\begin{itemize}
    \item In case of ($\texttt{D}^\texttt{H}$,$\texttt{A}^\texttt{H}$), all tracklets in $\Phi_{1:t}$ belong to CG1;
    \item In case of ($\texttt{D}^\texttt{L}$,$\texttt{A}^\texttt{H}$), a subset of tracklets in RoI belongs to CG1, and the rest of tracklets outside RoI belongs to CG3;
    \item In case of ($\texttt{D}^\texttt{H}$,$\texttt{A}^\texttt{L}$), all tracklets in $\Phi_{1:t}$ belong to CG2;
    \item In case of ($\texttt{D}^\texttt{L}$,$\texttt{A}^\texttt{L}$), a subset of tracklets in RoI belongs to CG2, and the rest of tracklets outside RoI belongs to CG3.
\end{itemize}
}

Tracklet confidence estimate $\overline{\Omega}(\chi_i^{t\text{+}1})$ at the $(t\text{+}1)$-th frame is then calculated by using Eqs.~(\ref{subeq:a})--(\ref{subeq:c}) in accordance with the category of $\chi_i^{t\text{+}1}$.
Finally, we define the \textit{expected} confidence score $\overline{\mathbf{\Omega}}_{\tau_k}(\Phi_{1:t\text{+}1},(\texttt{D}^x,\texttt{A}^y))$ of a MOT task $\tau_k$ for the choice of $(\texttt{D}^x$,$\texttt{A}^y)$, where $x$, $y \in \{ \texttt{H}, \texttt{L}\}$, at the $(t\text{+}1)$-th frame as 

\noindent
\begin{align}
    \overline{\mathbf{\Omega}}_{\tau_k}(\Phi_{1:t\text{+}1},(\texttt{D}^x,\texttt{A}^y)) = \frac{\sum_{\forall \chi_i^{t\text{+}1} \in \Phi_{1:t\text{+}1}} \overline{\Omega}(\chi_i^{t\text{+}1})}{|\Phi_{1:t\text{+}1}|}, \label{eq:expected_confidence}
\end{align}
where $|\Phi_{1:t\text{+}1}|$ is the total number of tracklets at the \mbox{$(t\text{+}1)$-th} frame.
We also define the \textit{measured} confidence score $\mathbf{\Omega}_{\tau_k}(\Phi_{1:t})$ of a task $\tau_k$ at the $t$-th frame as 

\noindent
\begin{align}
    \mathbf{\Omega}_{\tau_k}(\Phi_{1:t}) = \frac{\sum_{\forall \chi_i^{t} \in \Phi_{1:t}} \Omega(\chi_i^{t})}{|\Phi_{1:t}|}. \label{eq:measured_confidence}
\end{align}

Using the measure, 
the amount of \textit{expected} {increase} in $\tau_k$'s confidence score for the choice of $(\texttt{D}^x$,$\texttt{A}^y)$ at the $(t\text{+}1)$-th frame (denoted as $\Delta\overline{\mathbf{\Omega}}_{\tau_k}(\Phi_{1:t\text{+}1},(\texttt{D}^x,\texttt{A}^y))$) can be calculated as

\noindent
\begin{align}%\label{eq:delta_conf}
    \Delta\overline{\mathbf{\Omega}}_{\tau_k}(\Phi_{1:t\text{+}1},(\texttt{D}^x,\texttt{A}^y)) = \overline{\mathbf{\Omega}}_{\tau_k}(\Phi_{1:t\text{+}1},(\texttt{D}^x,\texttt{A}^y)) - \mathbf{\Omega}_{\tau_k}(\Phi_{1:t}).
    \label{eq:confidence_change}
\end{align}

{For simplicity of presentation, we will use the notation of $\Delta\overline{\mathbf{\Omega}}_{\tau_k}$, for $\Delta\overline{\mathbf{\Omega}}_{\tau_k}(\Phi_{1:t\text{+}1},(\texttt{D}^x,\texttt{A}^y))$.
By deriving $\Delta\overline{\mathbf{\Omega}}_{\tau_k}$,
we establish a reasonable criterion to determine which MOT task's frame should be executed among multiple MOT tasks, to be utilized by the scheduler.
That is, if we focus on improving overall tracking accuracy \textit{only},
the scheduler chooses the MOT task with the largest $\Delta\overline{\mathbf{\Omega}}_{\tau_k}$ (that yields the largest confidence score improvement). However, the scheduler also considers the timing guarantees of multiple MOT tasks,
to be addressed in the following section.}

%% file: 05scheduling.tex
\section{Scheduling Framework for \sys{}}
\label{sec:scheduling}

In this section, we consider 
\mbox{NPFP$^\textsf{min}$}, a traditional non-preemptive fixed-priority scheduling 
in which every instance (i.e., job) of a set of multi-object tracking tasks executes for the minimum execution requirement by selecting ($\texttt{D}^\texttt{L}$,$\texttt{A}^\texttt{L}$).
We then develop a novel scheduling framework designed for \sys{}, called \mbox{NPFP$^\textsf{flex}$} (Non-Preemptive Fixed-Priority with \textsf{flex}ible execution), so as to achieve two important design principles. 

\begin{itemize}
    \item First, \mbox{NPFP$^\textsf{flex}$} shares the existing offline schedulability test for \mbox{NPFP$^\textsf{min}$} (to be presented in Lemma~\ref{lemma:offline}),
    which offers timely execution of every instance (i.e. job) of a set of multi-object tracking tasks.
    \item Second, \mbox{NPFP$^\textsf{flex}$} checks the feasibility for each active job to be executed (regardless of its priority by FP) beyond
    its minimum execution requirement, without compromising the schedulability of any future jobs to executed according to \mbox{NPFP$^\textsf{min}$}. Among the active jobs that do not compromise the schedulability, \mbox{NPFP$^\textsf{flex}$} chooses to execute the job that yields 
    the largest expected improvement of confidence score among all pairs of a task and 
    ($\texttt{D}^{\texttt{L}/\texttt{H}}$,$\texttt{A}^{\texttt{L}/\texttt{H}}$)
    from Eq.~\eqref{eq:confidence_change},
    which in turn improves overall accuracy of MOT.
\end{itemize}

While it is difficult to embrace the second principle only, it is more challenging to establish the two principles together.
In this section, we address this real-time scheduling problem.

\subsection{Task Model}
\label{subsec:model}

To model multi-object tracking tasks that utilize \sys{}, 
we use a strictly periodic task model~\cite{LiLa73}.
A task $\tau_i\in\tau$ is specified by $(T_i,C_i)$,
where $T_i$ is the period and $C_i$ is the worst-case execution time (WCET).
A task $\tau_i$ invokes a series of jobs every $T_i$ times;
once a job is released at $t_0$, it should finish its execution no later than $t_0+T_i$.
A job is said to be \textit{active} at $t_0$, if it is released no later than $t_0$ and has remaining execution at $t_0$. 
Considering the non-preemptiveness of each job
(executed on GPU without preemption),
a job does not pause before completion, once it starts to execute.
We consider uniprocessor scheduling in which only a single job can be executed on the computing platform at any time; this accords with our system architecture in which every instance of a MOT task monopolizes GPU when it is executed.
 
The WCET
$C_i$ can be decomposed by $C_i^D$ and $C_i^A$, i.e., $C_i = C_i^D + C_i^A$.
The former is for the detection part, while the latter is for the association part.

In the detection part, $c_i^{infer}(\texttt{L}\text{ or }\texttt{H})$ is the total inference time for YOLOv5
that has the two 
different WCETs 
for low-confidence (\texttt{L}) and high-confidence (\texttt{H}) detection.
Then, $C_i^D$ can be calculated as follows: 
$C_i^{D}(\texttt{L}\text{ or }\texttt{H}) = c^{pre} + c_i^{infer}(\texttt{L}\text{ or }\texttt{H})$,
where $c^{pre}$ denotes the WCET for RoI identification and image cropping.

In the association part, $c_i^{as}(\texttt{L}\text{ or }\texttt{H})$ is the WCET
for low-confidence (\texttt{L}) and high-confidence (\texttt{H}) association.
As explained in Sec.~\ref{subsec:design},
the low-confidence association performs a simple IoU-based matching algorithm (whose WCET
is denoted by $c_i^{IoU}$),
while high-confidence association \textit{additionally} performs a feature-based method (whose WCET
is denoted by $c_i^{cascade}$). 
Therefore, $c_i^{as}(\texttt{L})$ = $c_i^{IoU}$ and $c_i^{as}(\texttt{H})$ = $c_i^{cascade} + c_i^{IoU}$ hold, which implies $c_i^{as}(\texttt{L}) < c_i^{as}(\texttt{H})$.
Then, we can express $C_i^A$ as follows:
$C_i^{A}(\texttt{L}\text{ or }\texttt{H}) =  c_i^{as}(\texttt{L}\text{ or }\texttt{H}) + c^{post}$,
where $c^{post}$ denotes the WCET for updating the confidence of all tracklets in each task.

Since we have two options (i.e., \texttt{L} and \texttt{H}) for each of the detection and association parts,
we have four options of $C_i$ (i.e., the WCET of $\tau_i$), where
the first and second characters  in the superscript of $C_i$ 
(each of which is either \texttt{L} or \texttt{H})
correspond to the detection and association parts, respectively. %, as follows.

\begin{itemize}%[leftmargin=10pt]
    \item $C_i^\texttt{LL} = C_i^D(\texttt{L})+C_i^A(\texttt{L})$: the WCET 
    of $\tau_i$ for ($\texttt{D}^\texttt{L}$,$\texttt{A}^\texttt{L}$)
    \item $C_i^\texttt{HL} = C_i^D(\texttt{H})+C_i^A(\texttt{L})$: the WCET 
    of $\tau_i$ for ($\texttt{D}^\texttt{H}$,$\texttt{A}^\texttt{L}$)
    \item $C_i^\texttt{LH} = C_i^D(\texttt{L})+C_i^A(\texttt{H})$: the WCET 
    of $\tau_i$ for ($\texttt{D}^\texttt{L}$,$\texttt{A}^\texttt{H}$)
    \item $C_i^\texttt{HH} = C_i^D(\texttt{H})+C_i^A(\texttt{H})$: the WCET 
    of $\tau_i$ for ($\texttt{D}^\texttt{H}$,$\texttt{A}^\texttt{H}$)
\end{itemize}

In this section, we employ FP (Fixed-Priority) scheduling in which each task has a static priority and each job inherits the priority of its invoking task.
Let $\textsf{HP}(\tau_i)$ and $\textsf{LP}(\tau_i)$ respectively denote a set of tasks whose priority is higher than $\tau_i$ and lower than $\tau_i$. 
The \textit{response time} of a job of $\tau_i$ is defined as the duration between the release and completion of the job.
A \textit{\mbox{level-$i$} busy period} is defined as the longest consecutive time interval during which the computing unit is occupied by jobs whose priority is higher than or equal to $\tau_i$. Let LHS and RHS denote the left-hand-side and right-hand-side, respectively.

\subsection{\mbox{NPFP$^\textsf{min}$}: Base Scheduling Algorithm}
\label{subsec:NPFPmin}

To develop \mbox{NPFP$^\textsf{flex}$}, we first explain its base scheduling algorithm \mbox{NPFP$^\textsf{min}$}, which is the same as the traditional non-preemptive fixed-priority scheduling with $C_i=C_i^\texttt{LL}$ for every $\tau_i\in\tau$.
We focus on $t_0$, at which at least one job is released when the computing platform is idle, or a job finishes its execution.
At every $t_0$, we choose the job of $\tau_i$, whose priority (inherited by its invoking task) is the highest among the jobs active at $t_0$.
We then execute the chosen job for ($\texttt{D}^\texttt{L}$,$\texttt{A}^\texttt{L}$)
during at most $C_i^\texttt{LL}$.

Then, we present an existing offline schedulability test for \mbox{NPFP$^\textsf{min}$} in the following lemma.

\vspace{5pt}

\begin{lemma}[In \cite{BTW95,YBB10,ChBr17a}] \label{lemma:offline}
Suppose that a task set $\tau$ is scheduled by the \mbox{NPFP$^\textsf{min}$} scheduling algorithm.
If every task $\tau_i\in\tau$ satisfies Eq.~\eqref{eq:rta1},
every job invoked by tasks in $\tau$ cannot miss its deadline.

    \begin{align}\label{eq:rta1}
    R_i \le T_i, 
    \end{align}
    \normalsize

    \noindent where $R_i$, the worst-case response time of $\tau_i$, is calculated by finding $R_i(x+1)=R_i(x)$ through iteration from $R_i(0)=C_i^\texttt{LL}+max_{\tau_j\in \textsf{LP}(\tau_i)} C_j^\texttt{LL}$ in Eq.~\eqref{eq:rta2}.

    \begin{align}\label{eq:rta2}
    R_i(x+1) = C_i^\texttt{LL}+\max_{\tau_j\in \textsf{LP}(\tau_i)} C_j^\texttt{LL} + \sum_{\tau_h\in\textsf{HP}(\tau_i)} \bigg\lceil\frac{R_i(x)}{T_h}\bigg\rceil\cdot C_h^\texttt{LL}
    \end{align}
    \normalsize
\end{lemma}
\begin{IEEEproof}
Without loss of generality, let $t=0$ be the release time of the job of $\tau_i$ (denoted by $J_i$).
Lemma~6 in \cite{GRS96} proves that the worst-case response time of $\tau_i$ (which is not necessarily from $J_i$, but can be from the following job of $\tau_i$) is found 
in a \mbox{level-$i$} busy period when all higher-priority jobs are released at $t=0$ and a lower-priority job whose execution time is the largest is released right before $t=0$.
In an interval of length $L$ that starts at $t=0$, the amount of execution of all jobs belonging to the former is upper-bounded by $\sum_{\tau_h\in\textsf{HP}(\tau_i)} \lceil\frac{L}{T_h}\rceil\cdot C_h^\texttt{LL}$, while the amount of the job that coincides with the latter is upper-bounded by $\max_{\tau_j\in \textsf{LP}(\tau_i)} C_j^\texttt{LL}$. Therefore, the RHS of Eq.~\eqref{eq:rta2} is the sum of the WCET of $J_i$ (i.e., $C_i^\texttt{LL}$) and that of all jobs executed before $J_i$'s execution. 
Therefore, if Eq.~\eqref{eq:rta1} holds for $\tau_i$,
the earliest job of $\tau_i$ (i.e., $J_i$) in any \mbox{level-$i$} busy period does not miss its deadline if it executes for up to $C_i^\texttt{LL}$.

In the second part of the proof, 
we prove $t_{x+1}-t_x \le T_i$,
where $t_x$ and $t_{x+1}$ respectively denote the time instants at which the $x^{th}$ and $(x+1)^{th}$ earliest jobs of $\tau_i$ start their execution
in the same \mbox{level-$i$} busy period;
the proof is similar to Lemma~2 of \cite{YBB10} that shows no self-pushing phenomenon under some conditions.
At $t_x$, there is no higher-priority active job; otherwise, the $x^{th}$ job cannot start its execution at $t_x$.
Therefore, in an interval of length $L$ that starts at $t_x$, the amount of executions of 
the $x^{th}$ job of $\tau_i$ and other jobs whose priority is higher than $\tau_i$ is $C_i^\texttt{LL}+\sum_{\tau_h\in\textsf{HP}(\tau_i)} \lceil\frac{L}{T_h}\rceil\cdot C_h^\texttt{LL}$  (denoted by $R_i'(L)$).
We can confirm that $R_i'(L)$ is no larger than the RHS of Eq.~\eqref{eq:rta2} with $R_i(x)=L$. Therefore, the supposition (i.e., $\tau_i$ satisfies Eq.~\eqref{eq:rta1}) implies that there exists $L=R_i'(L)$ that satisfies $R_i'(L)\le T_i$, meaning that the $(x+1)^{th}$ job of $\tau_i$ can start its execution no later than $T_i$ time units after $t_x$.

The first and second parts of the proof respectively operate as the base and inductive cases for mathematical induction, 
which proves the lemma.
\end{IEEEproof}

\subsection{\mbox{NPFP$^\textsf{flex}$}: Novel Scheduling Framework for \sys{}}
\label{subsec:NPFPflex}

We now develop \mbox{NPFP$^\textsf{flex}$} in Algorithm~\ref{algo:scheduling2}, designed for \sys{}.
In terms of the job prioritization and the job execution requirement,
\mbox{NPFP$^\textsf{flex}$} by default follows the policy of \mbox{NPFP$^\textsf{min}$} by setting \textsf{flex} to $\texttt{F}$ (Line~1), at every $t_0$ where at least a job is released when the computing platform is idle, or a job finishes its execution. 
Then, we check the feasibility for each active job (denoted by $J_k$ of $\tau_k$ in Line~2) to be executed for the given execution requirement $C_k$ (either $C_k^\texttt{LL}$, $C_k^\texttt{LH}$, $C_k^\texttt{HL}$, or $C_k^\texttt{HH}$ in Line~3).
To guarantee the feasibility of a job,
we need to guarantee (a) no deadline miss of $J_k$ if it starts its execution for $C_k$ at $t_0$,
\textit{and} (b) no deadline miss of all future jobs to be executed after $J_k$ according to \mbox{NPFP$^\textsf{min}$} (i.e., according to FP with $C_i=C_i^\texttt{LL}$);
(a) corresponds (i) in Line~4, while (b) corresponds to (ii) and (iii), to be explained in the later lemmas.
If it is deemed feasible to execute $J_k$ for given $C_k$, 
we calculate $\Delta\overline{\mathbf{\Omega}}_{\tau_k}$ in Eq.~\eqref{eq:confidence_change} (Line~5),
and update \textsf{flex} as \texttt{T} (Line~6).
After investigating the feasibility of all pairs of an active feasible job and its execution requirement (Lines 2--9),
we have two cases.
If $\textsf{flex}=\texttt{T}$, we choose to execute $J_k$ for $C_k$, whose $\Delta\overline{\mathbf{\Omega}}_{\tau_k}$ is the largest among all pairs of an active feasible job and given execution requirement (Lines~10--11).
Otherwise, we follow \mbox{NPFP$^\textsf{min}$}, implying we choose the highest-priority active job according to FP and execute the job during $C_i^\texttt{LL}$ for ($\texttt{D}^\texttt{L}$,$\texttt{A}^\texttt{L}$) (Lines~12--13).

\begin{algorithm} [t]
\caption{The \mbox{NPFP$^\textsf{flex}$} scheduling algorithm}
\label{algo:scheduling2}
\small
At $t_0$, at which at least a job is released when  the computing platform is idle, or a job finishes its execution,
\begin{algorithmic}[1]
    \STATE \textsf{flex} $\leftarrow$ \texttt{F}
    \FOR{Every active job
    (denoted by $J_k$ of $\tau_k$)}
        \FOR{$C_k \in \{C_k^\texttt{LL}, C_k^\texttt{LH}, C_k^\texttt{HL}, C_k^\texttt{HH}\}$, respectively for \{($\texttt{D}^{\texttt{L}}$,$\texttt{A}^{\texttt{L}}$),($\texttt{D}^{\texttt{L}}$,$\texttt{A}^{\texttt{H}}$),
        ($\texttt{D}^{\texttt{H}}$,$\texttt{A}^{\texttt{L}}$),
        ($\texttt{D}^{\texttt{H}}$,$\texttt{A}^{\texttt{H}}$)\}} 
            \IF {The following three conditions hold for assigned $C_k$: (i) Eq.~\eqref{eq:online_itself} holds, (ii) Eq.~\eqref{eq:online1} holds for all $\tau_j\in\tau$ with $Z_j(t_0)=\texttt{T}$, and (iii) Eq.~\eqref{eq:online2} holds for all $\tau_j\in\tau$ with $Z_j(t_0)=\texttt{F}$}
                \STATE Calculate $\Delta\overline{\mathbf{\Omega}}_{\tau_k}$ in Eq.~\eqref{eq:confidence_change} for given ($\texttt{D}^{\texttt{L}/\texttt{H}}$,$\texttt{A}^{\texttt{L}/\texttt{H}}$)
                \STATE \textsf{flex} $\leftarrow$ \texttt{T}
            \ENDIF
        \ENDFOR
    \ENDFOR
    \IF {\textsf{flex}=\texttt{T}}
        \STATE Execute $J_k$ for $C_k$, whose $\Delta\overline{\mathbf{\Omega}}_{\tau_k}$ is the largest.
    \ELSE
        \STATE Execute the job of $\tau_i$, whose priority is the highest among all the jobs active at $t_0$, during at most $C_i^\texttt{LL}$ for ($\texttt{D}^{\texttt{L}}$,$\texttt{A}^{\texttt{L}}$),
        which is the same as \mbox{NPFP$^\textsf{min}$}.
    \ENDIF
\end{algorithmic}
\normalsize
\end{algorithm}

We now present online feasibility tests for \mbox{NPFP$^\textsf{flex}$} in Line~4,
which are conditions for the job of interest $J_k$ 
and other jobs than $J_k$ not to miss their deadlines.
We then prove that \mbox{NPFP$^\textsf{flex}$} shares the same offline schedulability analysis for \mbox{NPFP$^\textsf{min}$} as Lemma~\ref{lemma:offline}.

For ease of presentation, we define two notions.
First, let $Z_i(t)$ denote the existence of an active job of $\tau_i$ at $t$; 
if $Z_i(t)=\texttt{T}$ and $Z_i(t)=\texttt{F}$, there exists an active job and no active job of $\tau_i$ at $t$, respectively.
Second, let $r_i(t)$ denote the earliest release time of any job of $\tau_i$ after or at $t$.
Since we consider the implicit-deadline periodic task model, 
$r_i(t)$ is not only a release time of the next job of $\tau_i$, 
but also an absolute deadline of the job of $\tau_i$ that is active at $t$ (if $Z_i(t)=\texttt{T}$).   

We focus on $t_0$ in Algorithm~\ref{algo:scheduling2}, at which at least one job is released when the computing platform
is idle, or a job finishes its execution under \mbox{NPFP$^\textsf{flex}$}.
Suppose that we start to execute an active job of $\tau_k$ (denoted by $J_k$) for given $C_k$ (either $C_k^\texttt{LL}$, $C_k^\texttt{LH}$, $C_k^\texttt{HL}$, or $C_k^\texttt{HH}$) at $t_0$.
$J_k$'s schedulability is simply checked in the following lemma.

\vspace{5pt}

\begin{lemma}\label{lemma:online_itself}
Suppose that we start to execute a job of $\tau_k$ (denoted by $J_k$) at $t_0$ for at most $C_k$.
Then, if Eq.~\eqref{eq:online_itself} holds, $J_k$ cannot miss its deadline.

    \begin{align}\label{eq:online_itself}
        C_k ~ \le ~ r_k(t_0)-t_0
    \end{align}
    \normalsize
\end{lemma}
\begin{IEEEproof}
By the non-preemptiveness, the execution time no larger than the time to its absolute deadline implies no deadline miss.
\end{IEEEproof}

\vspace{5pt}

Then, we check whether the earliest job of $\tau_j\in\tau$ to be executed after $J_k$'s execution is schedulable or not, with two cases: when there exists an active job of $\tau_j$ at $t_0$ (i.e., $Z_j(t_0)=\texttt{T}$), and no active job of $\tau_j$ at $t_0$ (i.e., $Z_j(t_0)=\texttt{F}$), respectively in Lemmas~\ref{lemma:online1} and \ref{lemma:online2}.

\vspace{5pt}

\vspace{5pt}

\begin{lemma}\label{lemma:online1}
Suppose that (i) we start to execute an active job of $\tau_k$ (denoted by $J_k$) at $t_0$ for at most $C_k$,
and (ii) all jobs to be executed after $J_k$'s execution are scheduled by \mbox{NPFP$^\textsf{min}$}.
If Eq.~\eqref{eq:online1} holds, the earliest job of a given $\tau_j$ with $Z_j(t_0)=\texttt{T}$ to be executed after $J_k$'s execution (denoted by $J_j$) cannot miss its deadline. Note that $\tau_j$ can be $\tau_k$.

    \begin{align}\label{eq:online1}
        & C_j^\texttt{LL} + C_k+
        \sum_{\tau_{h}\in\textsf{HP}(\tau_j)\setminus\{\tau_k\} | Z_{h}(t_0)=\texttt{T}}  C_{h}^\texttt{LL}  \nonumber\\
        & + \sum_{\tau_{h}\in\textsf{HP}(\tau_j) | %A_{h2}(t)=\texttt{L} ~\&~ 
        r_h(t_0)<r_j(t_0) } \bigg\lceil\frac{r_j(t_0)-r_h(t_0)}{T_h}\bigg\rceil \cdot C_h^\texttt{LL} \nonumber\\
        \le ~~~ & r_j(t_0)-t_0
    \end{align}
    \normalsize
\end{lemma}
\begin{IEEEproof}
Suppose that $J_j$ misses its deadline, even though Eq.~\eqref{eq:online1} holds.
Recall $r_j(t_0)$ is the absolute deadline of $J_j$ that is active at $t_0$.
Since the activeness of $J_j$ at $t_0$ and (ii) in the supposition of Lemma~\ref{lemma:online1}, a job whose priority is lower than $J_j$ (which is not $J_k$) cannot
be executed in $[t_0,r_j(t_0))$ before $J_j$'s execution.
Therefore, 
the only jobs that can execute before $J_j$'s execution in $[t_0,r_j(t_0))$
are (a) $J_k$, (b) all higher-priority jobs active at $t_0$ except $J_k$ (the ``except'' phrase is required only if $\tau_k\in\textsf{HP}(\tau_j)$), (c) all higher-priority jobs to be released after $t_0$. 
The WCET 
of (a) is $C_k$,
and that of (b) is the first summation term of the LHS of Eq.~\eqref{eq:online1}.
Considering $r_h(t_0)$ is the earliest job release time of given $\tau_h\in\textsf{HP}(\tau_j)$ after $t_0$,
the WCET
of (c) is upper-bounded by the second summation term of the LHS.
Therefore, missing $J_j$'s deadline implies that the sum of the WCET
of $J_j$ itself (i.e., $C_j^\texttt{LL})$, the WCET of (a),
the WCET of (b), and the WCET of (c) should be strictly larger than the interval length of $[t_0,r_j(t_0))$ (i.e., the RHS of Eq.~\eqref{eq:online1}). The sum is upper-bounded by the LHS of Eq.~\eqref{eq:online1},
which contradicts Eq.~\eqref{eq:online1}.
Therefore, $J_j$ cannot miss its deadline if it executes for up to $C_j^\texttt{LL}$.
\end{IEEEproof}

\vspace{5pt}

\begin{lemma}\label{lemma:online2}
Suppose that (i) we start to execute an active job of $\tau_k$ (denoted by $J_k$) at $t_0$ for at most $C_k$,
(ii) all jobs to be executed after $J_k$'s execution are scheduled by \mbox{NPFP$^\textsf{min}$}, and
(iii) Eq.~\eqref{eq:rta1} holds for every $\tau_i\in\tau$.
If Eq.~\eqref{eq:online2} holds, the earliest job of a given $\tau_j$ with $Z_j(t_0)=\texttt{F}$ to be executed after $J_k$'s execution (denoted by $J_j$) cannot miss its deadline.
Note that $\tau_j$ cannot be $\tau_k$, as the existence of $J_k$ implies $Z_k(t_0)=\texttt{T}$.

    \begin{align}\label{eq:online2}
        & C_j^\texttt{LL} + C_k+
        \sum_{\tau_{h}\in\textsf{HP}(\tau_j)\setminus\{\tau_k\} | Z_{h}(t_0)=\texttt{T}} C_{h}^\texttt{LL}  \nonumber\\
        & + \sum_{\tau_{h}\in\textsf{HP}(\tau_j) | %A_{h2}(t)=\texttt{L} ~\&~ 
        r_h(t_0)<r_j(t_0)+T_j } \bigg\lceil\frac{r_j(t_0)+T_j-r_h(t_0)}{T_h}\bigg\rceil \cdot C_h^\texttt{LL} \nonumber\\
        \le ~ & r_j(t_0)+T_j-t_0
    \end{align}
    \normalsize
\end{lemma}
\begin{IEEEproof}
Suppose that $J_j$ misses its deadline, even though Eq.~\eqref{eq:online2} holds.
Recall $r_j(t_0)+T_j$ is the absolute deadline of $J_j$ because $J_j$ is not active at $t_0$ implying $r_j(t_0)$ is the release time of $J_j$.
We consider two cases.

(Case 1) We consider the case where the computing platform is idle or occupied by a job that is not $J_k$ but has lower priority than $J_j$. Let $t'$ denote the latest time instant belongings to the case.
By the definition of $t'$, the interval from $t'$ to the end of $J_j$'s execution is included in a \mbox{level-$j$} busy period that starts from $t'$.
By (ii) in the supposition of Lemma~\ref{lemma:online2},
all jobs executed after $J_k$'s execution are scheduled by \mbox{NPFP$^\textsf{min}$}.
On the other hand, the proof of Lemma~\ref{lemma:offline} shows that if Eq.~\eqref{eq:rta1} holds for $\tau_j\in\tau$, any job of $\tau_j$ in any \mbox{level-$j$} busy period 
does not miss its deadline when $\tau$ is scheduled by \mbox{NPFP$^\textsf{min}$}.
Therefore, if we consider the proof of Lemma~\ref{lemma:offline},
$J_j$'s deadline miss contradicts (iii) in the supposition of Lemma~\ref{lemma:online2}.

(Case 2: the opposite of Case 1)
In this case, the proof is the same as that of Lemma~\ref{lemma:online1} by changing the interval of interest from $[t_0,r_j(t_0))$ to $[t_0,r_j(t_0)+T_j)$.

By Cases 1 and 2, $J_j$ cannot miss its deadline
if it executes for up to $C_j^\texttt{LL}$.
\end{IEEEproof}

\vspace{5pt}

While Lemmas~\ref{lemma:online1} and \ref{lemma:online2} guarantee the schedulability of the earliest job of each $\tau_j$ to be executed after $J_k$'s execution, we need to guarantee the schedulability of every job of $\tau_j$ to be executed after $J_k$'s execution, as follows.

\vspace{5pt}

\begin{lemma}\label{lemma:next_job}
Suppose that (i) a job of $\tau_j$ (denoted by $J_j$) 
starts its execution at $t_1$
and does not miss its deadline if it executes for up to $C_j^\texttt{LL}$,
ted after $J_j$'s execution are scheduled by \mbox{NPFP$^\textsf{min}$}, and
(iii) Eq.~\eqref{eq:rta1} holds for every $\tau_i\in\tau$.
Then, any job of $\tau_j$ to be executed after $J_j$'s execution cannot miss its deadline.
\end{lemma}
\begin{IEEEproof}
Let $J_j'$ denote the next job of $J_j$ invoked by $\tau_j$.
We focus on
$[t_1,t_2)$, 
where $t_2$ is the time instant at which $J_j'$
starts its execution.
We prove no deadline miss of $J_j'$. 

(Case 1: no processor idle status and no execution of jobs whose priority is lower than $\tau_j$ in $[t_1,t_2)$)
Since $J_j$ and $J_j'$ are executed in the same \mbox{level-$j$} busy period,
we can prove that $t_2-t_1\le T_j$, by applying the technique in the second part of the proof of Lemma~\ref{lemma:offline} (corresponding to $t_{x+1}-t_x\le T_i$).
Therefore, (i) in the supposition of Lemma~\ref{lemma:next_job}
implies
no deadline miss of $J_j'$ if it executes up to $C_j^\texttt{LL}$ (which is guaranteed by (ii) in the supposition of Lemma~\ref{lemma:next_job}).

(Case 2: the opposite case of Case 1)
Let $t'$ denote the latest time instant in $[t_1,t_2)$ %$[t_1+X,t_2+X')$
at which the computing platform is idle or occupied by a job whose priority is lower than $J_j'$. 
The remaining proof is the same as Case 1 of the proof of Lemma~\ref{lemma:online2}.

By Cases 1 and 2, $J_j'$ cannot miss its deadline. % if it executes up to $C_j^\texttt{LL}$.
Applying the same technique as proving no deadline miss of $J_j'$ from no deadline miss of $J_j$,
we can prove that all following jobs of $\tau_j$ do not miss their deadlines.
\end{IEEEproof}

\vspace{5pt}

Utilizing Lemmas~\ref{lemma:online_itself}, \ref{lemma:online1}, \ref{lemma:online2}, and \ref{lemma:next_job},
we develop an offline schedulability analysis for \mbox{NPFP$^\textsf{flex}$}, which is the same as that for \mbox{NPFP$^\textsf{min}$} (i.e., Lemma~\ref{lemma:offline}).

\vspace{5pt}

\begin{theorem}\label{theorem:NPFP_FLEX}
Suppose that a task set $\tau$ is scheduled by \mbox{NPFP$^\textsf{flex}$} in Algorithm~\ref{algo:scheduling2}.
If every task $\tau_i\in\tau$ satisfies Eq.~\eqref{eq:rta1},
every job invoked by tasks in $\tau$ cannot miss its deadline.
\end{theorem}
\begin{IEEEproof}
Suppose that a job misses its deadline at $t_a$. Then, there should exists the latest time instant $t_0$ before $t_a$, at which a job of $\tau_k$ %$\in\tau$ 
(denoted by $J_k$) starts its execution although it is not the highest-priority active job and/or it executes for more than $C_k^\texttt{LL}$.
Otherwise, until $t_a$, all jobs are scheduled according to \mbox{NPFP$^\textsf{min}$}; the existence of a job deadline miss contradicts Lemma~\ref{lemma:offline}.

By the definition of $t_0$, in the time interval from the end of $J_k$'s execution to $t_a$,
all jobs are scheduled according to \mbox{NPFP$^\textsf{min}$}. 
Also, by the policy of \mbox{NPFP$^\textsf{flex}$}, $J_k$ starts its execution at $t_0$ only if 
the following three conditions hold for assigned $C_k$ (in Line~4 of Algorithm~\ref{algo:scheduling2}): (i) Eq.~\eqref{eq:online_itself} holds, (ii) Eq.~\eqref{eq:online1} holds for all $\tau_j\in\tau$ with $Z_j(t_0)=\texttt{T}$, and (iii) Eq.~\eqref{eq:online2} holds for all $\tau_j\in\tau$ with $Z_j(t_0)=\texttt{F}$.
(i) implies $J_k$ cannot miss its deadline by Lemma~\ref{lemma:online_itself},
(ii) and (iii) imply the earliest job of every $\tau_j$ to be executed after $J_k$'s execution cannot miss its deadline by Lemmas~\ref{lemma:online1} and \ref{lemma:online2}.
By Lemma~\ref{lemma:next_job}, (ii) and (iii) imply every job of $\tau_j$ to be executed after $J_k$'s execution cannot miss its deadline.
Therefore, the existence of a job deadline miss at $t_a$ contradicts either Lemma~\ref{lemma:online_itself}, \ref{lemma:online1}, \ref{lemma:online2}, or \ref{lemma:next_job}, which proves the theorem.
\end{IEEEproof}

\vspace{5pt}

\textbf{Run-time complexity of \mbox{NPFP$^\textsf{flex}$}.}
At each $t_0$ (at which a job is released or completed in Algorithm~\ref{algo:scheduling2}),
we need to perform Line~4 of Algorithm~\ref{algo:scheduling2} for every active job $J_k$.
Since checking Eq.~\eqref{eq:online1} or \eqref{eq:online2} takes $O(n)$ for given $J_k$ and $\tau_j$,
it takes $O(n^2)$ to perform Line~4 of Algorithm~\ref{algo:scheduling2} for given $J_k$,
where $n$ is the number of tasks in $\tau$.
Therefore, at each $t_0$, \mbox{NPFP$^\textsf{flex}$} requires $O(n^2\cdot n')$, 
where $n'$ is the number of active jobs at $t_0$.
Note that the number of cameras of an autonomous vehicle (i.e., $n$) is only a few,
e.g., one each in the front, rear and both sides of the vehicle, and one each between the front and both sides; therefore, \sys{} provides an acceptable scheduling overhead with $O(n^2\cdot n')$ time-complexity, to be demonstrated with experimental results in Sec. VI-A.

%% file: 06evaluation.tex
\section{Evaluation}
\label{sec:evaluation}

In this section, we first explain our experimental setup. We then present experimental results of \sys{}, compared to existing methods.

\subsection{Experimental Setup}

\textbf{Hardware and software.}
We conduct experiments on a computer equipped with an Intel(R) Xeon(R) Silver 4215R CPU @ 3.20GHz, 251.5GB RAM, and NVIDIA V100 GPU. 
We use Ubuntu 18.04.4 with CUDA 10.2, and PyTorch 1.10.2 for implementation. 
We employ YOLOv5 as a front-end detector with the different input sizes of 672$\times$ 672 and 256$\times$ 256, for high- and low-confidence detection, respectively. 
We also employ SORT~\cite{sort} and DeepSORT~\cite{deepsort} with the re-identification model of OSNet~\cite{zhou2019omni} for high- and low-confidence association, respectively.
We use the Waymo Open Dataset~\cite{waymo} in our evaluation. 

\textbf{Execution time profiling and run-time overhead.}
\sys{} conducts DNN-based computations on a GPU, while others including IoU matching and confidence estimation use
a CPU.
We take a measurement-based approach to estimate the worst-case execution time of detection and association parts for each multi-object tracking task, $C_i^{D}$ and $C_i^{A}$, under \sys{}.
Our experiments measure the execution time of each component in $C_i^{D}$ and $C_i^{A}$ by running it 1,000 times, taking the maximum value among the measured ones as the WCET, % as 
summarized in Table~\ref{table1}.
For example, the WCETs for low- and high-confidence detection ($c_{i}^{infer}(\texttt{L})$ and $c_{i}^{infer}(\texttt{H})$) are 17.6ms and 23.2ms, while the WCETs for low- and high-confidence association ($c_{i}^{as}(\texttt{L})$ and $c_{i}^{as}(\texttt{H})$) are 9.6ms and 32.7ms, respectively.
The two key modules for RoI extraction and confidence estimation ($c^{pre}$ and $c^{post}$) in \sys{} take 0.6ms and 0.7ms on average, respectively, a relatively short delay compared to the execution times for detection and association ($c_i^{infer}$ and $c_i^{as}$).

We also measure the run-time overhead of our scheduling framework for \sys{} as a function of the number of MOT tasks ($n$).
When $n$ is increased from 1 to 10, the run-time scheduling overhead increases from 0.5ms to 0.7ms, which is almost linear in $n$ with low slope.

\subsection{Experimental Results}
We would like to demonstrate the capability of \sys{} making a significant improvement in multi-object tracking accuracy while meeting all timing requirements for multiple MOT tasks. 
We use MOTA~\cite{mota}, a primary metric to evaluate the accuracy of multi-object tracking algorithms.
The MOTA metric deals with both detection and tracker outputs by taking into consideration of miss detection, false detection, and false tracking. 
Note that a higher MOTA score indicates higher overall tracking accuracy.

\begin{table}
\setlength{\tabcolsep}{3pt}.
\centering
\caption{Execution time measurement for each component}
\begin{tabular}{c|ccc|ccc} 
\hline
\multirow{2}{*}{Time(ms)} & \multicolumn{3}{c|}{$C_{i}^{D}$}                                & \multicolumn{3}{c}{$C_{i}^{A}$}                                       \\
                          & \multicolumn{1}{l}{$c^{pre}$} & $c_{i}^{infer}(\texttt{L})$ & $c_{i}^{infer}(\texttt{H})$ & $c_{i}^{as}(\texttt{L})$ & $c_{i}^{as}(\texttt{H})$ & \multicolumn{1}{l}{$c^{post}$}  \\ 
\hline
Average                   & 0.6                           & 12.6           & 13.1           & 3.2            & 23.4           & 0.7                                 \\
Maximum                   & 0.9                           & 17.6           & 23.2           & 9.6            & 32.7           & 0.9                                 \\
\hline
\end{tabular}
\label{table1}
\end{table}

We evaluate three versions of \sys{}, 
which (i) employ different job prioritization and execution requirements for detection and association, but (ii) share the same offline schedulability analysis of Eq.~\eqref{eq:rta1} in Lemma~\ref{lemma:offline}.

\begin{itemize}
    \item \mbox{\sys{}$^\textsf{min}$} employs \mbox{NPFP$^\textsf{min}$} (i.e., always performing \texttt{L} detection and \texttt{L} association).
    \item \mbox{\sys{}$^\textsf{flex-NPI}$} employs \mbox{NPFP$^\textsf{flex}$} but does not allow priority inversion, i.e., the feasibility of $J_k$ for \texttt{H} detection and/or \texttt{H} association in Lines 3--8 of Algorithm~\ref{algo:scheduling2} is tested only for the job whose priority is the highest among all active jobs rather than every active job in Line~2.
    \item \mbox{\sys{}$^\textsf{flex}$}: \sys{} employs \mbox{NPFP$^\textsf{flex}$} in Algorithm~\ref{algo:scheduling2} as it is.
\end{itemize}

\noindent
As a prioritization policy for FP, we employ RM (Rate Monotonic).

We compare the \sys{} versions with two existing popular multi-object tracking approaches.

\begin{itemize}
    \item \mbox{\small\textsf{H+SORT}} employs YOLOv5 with the original frame size (672$\times$672) for detection and SORT for association (that corresponds to \texttt{L} association of \sys{}).
    \item \mbox{\small\textsf{L+DeepSORT}} employs YOLOv5 with the down-scaled frame size (256$\times$256) for detection and DeepSORT for association (that corresponds to \texttt{H} association of \sys{}).
    \item \mbox{\small\textsf{H+DeepSORT}} employs YOLOv5 with the original frame size (672$\times$672) for detection and DeepSORT for association.
\end{itemize}

\noindent 
Note that we exclude \mbox{\small\textsf{L+SORT}} for comparison because it shows a similar result to \mbox{\sys{}$^\textsf{min}$}.
For a fair comparison, 
\mbox{\small\textsf{H+SORT}}, \mbox{\small\textsf{L+DeepSORT}}, and \mbox{\small\textsf{H+DeepSORT}} employ non-preemptive fixed-priority scheduling with the same offline schedulability analysis of Eq.~\eqref{eq:rta1}. 

\begin{figure}[t]
	\centering
	\includegraphics[width=1\linewidth]{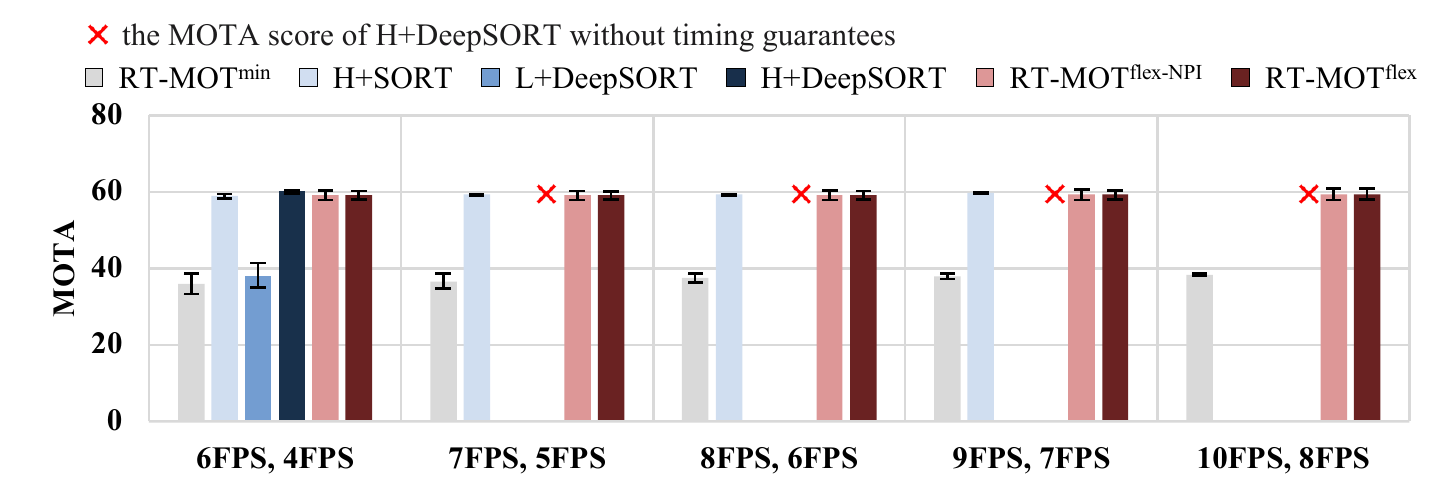}
	\caption{Tracking accuracy comparison for task sets with two tasks}
	\label{fig:eval1}
\end{figure} 

Fig.~\ref{fig:eval1} compares the average MOTA scores (plotted as bar graphs) and the maximum and minimum MOTA scores (plotted as error bars)
under six different approaches for five different task sets. 
Each of the task sets consists of two tasks with different FPS requirements, from the smallest workload
(6FPS and 4FPS) in the left-most, to the largest one (10FPS and 8FPS) in the right-most,
increased by 1FPS each for both tasks.
Note that, if a task set is not deemed schedulable by the offline schedulability analysis in Eq.~\eqref{eq:rta1} under the target MOT approach, we do not include to plot MOTA scores as we cannot offer an offline timing guarantee.
We observe that \mbox{\sys{}$^\textsf{flex-NPI}$} and \mbox{\sys{}$^\textsf{flex}$} consistently show much higher overall accuracy than \mbox{\sys{}$^\textsf{min}$} across all task sets (1.5$\times$ accuracy improvement). 
\mbox{\small\textsf{H+DeepSORT}} shows the highest overall accuracy (with a marginal improvement over \mbox{\sys{}$^\textsf{flex}$} by 1.0\%p) for the task set with 6FPS and 4FPS, but it cannot achieve timely execution for other task sets due to heavy computational workloads.
\mbox{\small\textsf{H+SORT}} also shows a comparable accuracy with \mbox{\sys{}$^\textsf{flex}$} for three task sets with up to 9FPS and 7FPS, but it cannot achieve timely execution for the task set with 10FPS and 8FPS. 

Fig.~\ref{fig:eval2} shows the accuracy results for three task sets, each of which consists of four tasks.
In general, we observe a similar trend to Fig.~\ref{fig:eval1};
all MOT approaches except \mbox{\sys{}$^\textsf{min}$} and \mbox{\small\textsf{L+DeepSORT}}
exhibit comparable MOTA scores under the smallest workload (8FPS, 4FPS, 2FPS, and 1FPS),
while only \mbox{\sys{}$^\textsf{flex-NPI}$} and \mbox{\sys{}$^\textsf{flex}$} perform well under the largest workload (10FPS, 6FPS, 4FPS, and 3FPS).
However, for the task set with the largest workload,
\mbox{\sys{}$^\textsf{flex}$} exhibits a smaller difference in accuracy among tasks by only up to 4.4\%p, while \mbox{\sys{}$^\textsf{flex-NPI}$} exhibits a larger difference by up to 13\%p;
also, we observe 3.1\%p accuracy degradation of \mbox{\sys{}$^\textsf{flex-NPI}$},
compared to \mbox{\sys{}$^\textsf{flex}$}.
Such a result can be interpreted as the benefit of \mbox{\sys{}$^\textsf{flex}$} that enables flexible job-level scheduling by allowing bounded priority inversions without violating any FPS requirement.
Although \mbox{\sys{}$^\textsf{min}$}, \mbox{\small\textsf{H+SORT}}, \mbox{\small\textsf{L+DeepSORT}}, and \mbox{\small\textsf{H+DeepSORT}} are  \emph{static} approaches that apply a fixed choice of detection and association models across all frames, \mbox{\sys{}$^\textsf{flex-NPI}$} and \mbox{\sys{}$^\textsf{flex}$} dynamically determine a pair of detection and association models frame-by-frame by considering both available resources and expected confidence improvement at run-time, achieving higher overall tracking accuracy while satisfying all FPS requirements. 
For example, for the task set with the largest workload,
\mbox{\sys{}$^\textsf{flex}$} flexibly selects ($\texttt{D}^{\texttt{L}}$,$\texttt{A}^{\texttt{L}}$), ($\texttt{D}^{\texttt{H}}$,$\texttt{A}^{\texttt{L}}$), and ($\texttt{D}^{\texttt{H}}$,$\texttt{A}^{\texttt{H}}$) for 52, 70, and 336 frames, respectively, out of 458 total jobs. 
Note that ($\texttt{D}^{\texttt{L}}$,$\texttt{A}^{\texttt{H}}$) is rarely selected in our experiment because i) $c_{i}^{as}(\texttt{H})$ is larger than $c_{i}^{infer}(\texttt{H})$ by 1.4$\times$ as observed in Table~\ref{table1}, and ii) \mbox{\small\textsf{L+DeepSORT}} shows a marginal accuracy improvement over \mbox{\sys{}$^\textsf{min}$} as observed in Figs.~\ref{fig:eval1} and~\ref{fig:eval2}.
Also, note that the maximum achievable MOTA score by \mbox{\small\textsf{H+DeepSORT}} \emph{without timing guarantees} is 60 (marked as red cross) for the task set with the largest workload, while the MOTA score of \mbox{\sys{}$^\textsf{flex}$} is 59.1. 
Therefore, \mbox{\sys{}$^\textsf{flex}$} can achieve nearly maximum tracking accuracy with $0.84\times$ less total computation time as compared to \mbox{\small\textsf{H+DeepSORT}}, making the task set schedulable.

In summary, \mbox{\sys{}$^\textsf{flex}$} can be adapted to various task sets through the dynamic selection of a pair of detection and association execution models and flexible scheduling frame-by-frame, achieving high overall tracking accuracy while meeting all FPS requirements at run-time in addition to offline timing guarantees. 

\vspace{-0.15cm}

%% file: 07related_work.tex
\begin{figure}[t]
	\centering 
	\includegraphics[width=1\linewidth]{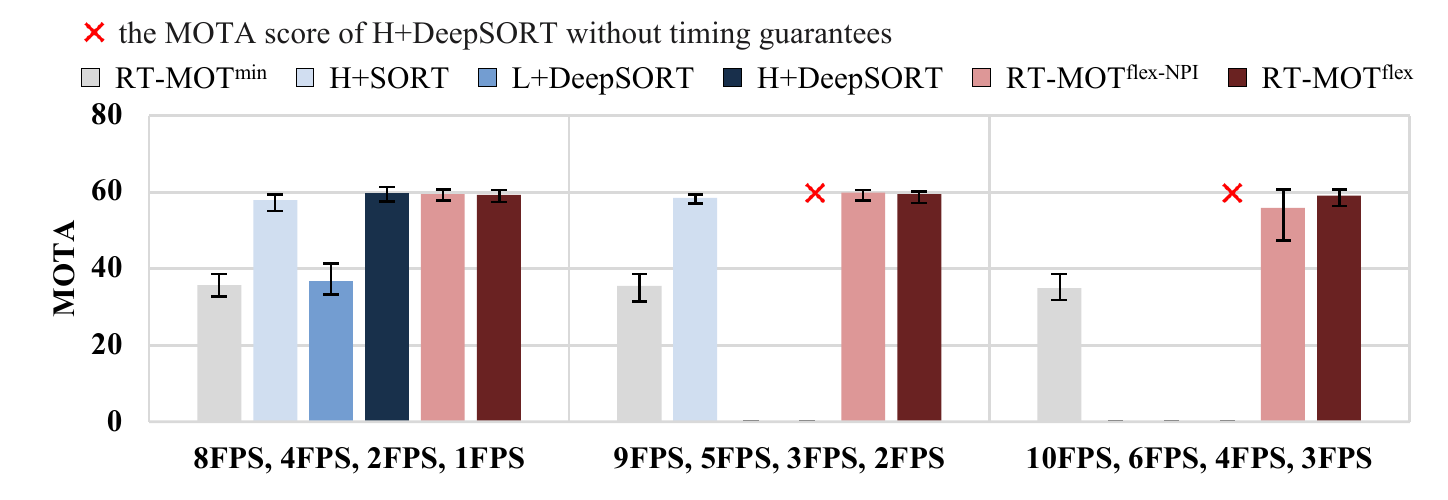}
	
	\caption{Tracking accuracy comparison for task sets with four tasks} 
	\label{fig:eval2}
\end{figure}

\section{Related Work}
\label{sec:related_work}

MOT can be categorized into one-stage and two-stage models depending on where in detection and association the AI (Artificial Intelligence) model is used. \sys{} uses two-stage MOT algorithms such as DeepSORT~\cite{deepsort} and StrongSORT~\cite{strongsort}, which employ two different models each for both detection and association.
Two-stage MOT uses the detection model to extract the detected object with the re-identification model and then associates it with matching algorithms. 
On the other hand, a one-stage MOT only uses a single model during detection and association. Objects and their features can be found within a single inference, and much like a two-stage MOT, the association process utilizes matching algorithms, e.g., FairMOT~\cite{fairmot} and BytesTrack~\cite{bytetrack}.

While most one-stage and two-stage models for MOT have focused on high tracking accuracy and average FPS, several studies have attempted to address both timing guarantee and high detection accuracy for the object detection~\cite{KAO18, KBS20, LeNi20, HCK20}, which is one of the main parts of two-stage MOT.
Since those studies  do not take the execution for association into consideration, 
their techniques for timing guarantee cannot be directly applied to MOT. 
When it comes to MOT itself, 
there has been a sole study that has achieved both timing guarantee and high tracking accuracy~\cite{self-cueing}.
By defining and utilizing the notion of uncertainty, the study in~\cite{self-cueing} has achieved a timing guarantee of the scheduling horizon while improving location accuracy.
However, since the study has targeted a single MOT task, it cannot be applied to multiple MOT tasks, which is different from \sys{}.

%% file: 08conclusion.tex
\section{Conclusion}
\label{sec:conclusion}

In this paper, we proposed a novel confidence-aware real-time scheduling framework 
for multiple MOT tasks, \sys{}, which consists of 
(i) a method to estimate the overall accuracy variation according to different detector/tracker selections, (ii) a scheduling framework that provides offline timing guarantees while maximizing overall accuracy at run-time using the method,
and (iii) a system architecture that supports the framework.
Through experiments, we demonstrated that \sys{} 
achieves high accuracy and timely execution of a set of MOT tasks, which has not been accomplished by existing tracking-by-detection methods.
In the future, we would like to extend our framework towards more combinations of various detectors and trackers.
In addition, we would like to improve the scheduling framework in terms of schedulability performance, e.g., by allowing a preemption between the completion of detection and the beginning of association for each MOT task.